\documentclass{elsart}
\usepackage[]{latexsym,amsmath,amssymb}
\usepackage[]{epsfig}

\def\simgt{\stackrel{>}{{}_\sim}}

\begin{document}
\begin{frontmatter}
\begin{flushright}
\texttt{DESY 08-033\\IZTECH-P-08/02}
\end{flushright}%
\title{\bf Inflation with Non-Minimal Coupling:\\ Metric vs. Palatini Formulations}
\author[DESY]{Florian Bauer}
\ead{florian.bauer@desy.de}
\and 
\author[DESY,IZMIR]{Durmu{\c s} A. Demir}
\ead{durmus.demir@desy.de}
\address[DESY]{Deutsches Elektronen-Synchrotron, DESY, 22607 Hamburg, Germany}
\address[IZMIR]{Department of Physics, Izmir Institute of Technology, TR35430, Izmir, Turkey}
%
%
%
%
\begin{abstract}
We analyze non-minimally coupled scalar field theories in metric (second-order) and Palatini (first-order) formalisms in a comparative fashion. After contrasting them in a general setup, we specialize to inflation and find that the two formalisms differ in their predictions for various cosmological parameters. The main reason is that dependencies on the non-minimal coupling parameter are different in the two formalisms. For successful inflation, the Palatini approach prefers a much larger value for the non-minimal coupling parameter than the Metric approach. Unlike the Metric formalism, in Palatini, the inflaton stays well below the Planck scale whereby providing a natural inflationary epoch.
\end{abstract}
\end{frontmatter}
\section{Introduction}
Scalar fields play a fundamental role in various physical
phenomena ranging from electroweak symmetry breaking to cosmological
inflation. Though no fundamental scalar
has been observed to date, they are an indispensable part of the
theoretical landscape: They can trigger breakdown of gauge
symmetries to induce particle masses, they can facilitate
inflation to explain shortcomings of standard cosmology, they can
make up quintessence to explain amount of dark energy in the
universe, and they can act as source of the non-baryonic dark
matter in the universe.

Given this rather widespread role of scalar fields at various
energy scales, it is of physical relevance to examine their dynamics
in geometries more general than general relativity (based
on {\it Metric formulation} or second-order formalism). Indeed, huge
discrepancy between experimental and quantum-theoretic values of
the dark energy, absence of any direct observation of dark matter
particles though a considerable amount is needed for cosmological
concordance, and unknown nature of gravitational dynamics at small
and large distances all motivate consideration of a more
general geometrical arena than general relativity. Given that there is no
unique way of generalization, one possibility, among many one can
consider, is to exploit certain features of the affine spaces. The
essential observation is that the affine connection and metric tensor
are a priori independent geometrical variables, and if they
are to exhibit any relationship it must arise from dynamical
equations a posteriori. This very setup, the so-called
{\it Palatini formulation} or first-order formalism \cite{palatini}, does not necessarily
admit the Levi-Civita connection if the matter sector depends
explicitly on the affine connection \cite{differ,differ2}.

This work is devoted to a comparative analysis of the dynamics of
non-minimally coupled scalar fields in Metric and Palatini formulations.
It will be shown that the two formalisms generally yield different
answers for both metric tensor and scalar fields provided that
non-minimal coupling of the scalar field to curvature scalar does not
vanish. The analysis will be targeting, for definiteness and concreteness,
scalar field theories which can facilitate both electroweak symmetry breaking 
and inflation.

In the next section we give a comparative discussion of Palatini and
Metric formalisms in a general setting in which a scalar field
interacts non-minimally with the spacetime curvature. In Sec. 3 we analyze
inflation with non-minimally coupled scalar field, and contrast predictions
of the two formalisms by computing various inflationary parameters. In
Sec. 4 we conclude.

\section{Metric vs. Palatini Formulations: Generalities}
In this section, we give a brief discussion of the Metric and
Palatini formulations in a comparative fashion within a
general setup.

An affine connection, whose components to be symbolized by ${\Gamma}^{\lambda}_{\alpha
\beta}$, governs parallel transport of tensor fields along a given
curve in spacetime, and parallel transport around a closed curve,
after one complete cycle, results in a finite mismatch if the
spacetime is curved. This curving is uniquely determined by the
Riemann tensor $\mathbb{R}^{\mu}_{\alpha \nu \beta}\left(
\Gamma\right)$ whose contraction $\mathbb{R}_{\alpha \beta}\left(
\Gamma\right) \equiv \mathbb{R}^{\mu}_{\alpha \mu \beta}\left(
\Gamma\right)$ gives the Ricci tensor\footnote{
The affine connection determines not only the curving but also the
twirling of the spacetime. The latter is encoded in the torsion
tensor \cite{differ,torsion}, and it is discarded from the analysis here by
considering a symmetric connection $\Gamma^{\lambda}_{\alpha \beta} =
\Gamma^{\lambda}_{\beta \alpha}$ only.}.

The spacetime gets further structured by the notion of distance if
it is endowed with a metric tensor $g_{\mu \nu}$ representing
clocks and rulers. The connection coefficients and metric tensor
are fundamentally independent quantities. They exhibit no {\it a
priori} known relationship, and if they are to have any it must
derive from a additional constraint or geometrodynamics \cite{palatini}.

For explicating differences and similarities between the Metric and
Palatini approaches it proves useful to focus on a generic action
\begin{eqnarray}
\label{action} \mathcal{S} = \int d^{4} x\,  \sqrt{-g} \left\{- \frac{1}{2}
{M}^2 g^{\mu \nu} \mathbb{R}_{\mu\nu}\left(\Gamma \right) +
{\mathcal{L}}_{\text{m}}\left(\varphi, \Gamma, g\right)\right\}
\end{eqnarray}
comprising gravitational (taken to have Einstein-Hilbert form) and
material (parameterized by the Lagrangian
${\mathcal{L}}_{\text{m}}\left(\varphi, \Gamma, g\right)$ involving the
matter fields $\varphi$, metric tensor $g_{\alpha \beta}$, and
possibly also the affine connection $\Gamma^{\lambda}_{\alpha
\beta}$) sectors. The mass parameter $M$ may or may not be
identical to the fundamental scale of gravity $M_{\text{Pl}}=
\left(8 \pi G_N\right)^{-1/2}$ since ${\mathcal{L}}_{\text{m}}\left(\varphi,
\Gamma, g\right)$ can feed further contributions to Newton's
constant.

The Metric formulation refers to replacing $\Gamma$ in (\ref{action}) by $\overline{\Gamma}$ where
\begin{eqnarray}
\label{levi}
\overline{\Gamma}^{\lambda}_{\alpha \beta} = \frac{1}{2} g^{\lambda \rho} \left( \partial_{\alpha} g_{\beta \rho}
+ \partial_{\beta} g_{\rho \alpha} - \partial_{\rho} g_{\alpha \beta}\right)
\end{eqnarray}
is the Levi-Civita connection. Having this constraint, the metric tensor $g_{\alpha \beta}$ and matter
fields $\varphi$ remain as the only independent dynamical variables, and
extremization of the action (\ref{action}) with respect to them yields the
associated equations of motion. One notes, however, that the curvature tensor
involves second derivatives of the metric tensor, and thus, cancellation of
the associated surface terms requires enhancement of (\ref{action}) by an extrinsic curvature contribution.

The Palatini formulation refers to keeping the metric, the connection and matter fields as independent dynamical variables in (\ref{action}). Therefore, the affine connection is not a predetermined quantity at all; it takes the form that dynamics requires. Moreover, the action involves only the first derivatives of the connection, and hence, there arises no surface term to be cancelled by an extrinsic curvature contribution.

An important feature of the Palatini formulation is that if ${\mathcal{L}}_{\text{m}}\left(\varphi, \Gamma, g\right)$ is
independent of $\Gamma^{\lambda}_{\alpha \beta}$ then the equation of motion for the connection
\begin{eqnarray}
\label{relax}
\nabla^{\Gamma}_{\lambda}\left(\sqrt{-g} g^{\alpha \beta}\right)=0
\end{eqnarray}
uniquely returns $\Gamma = \overline{\Gamma}$. Therefore, the Levi-Civita connection arises dynamically with no need to extrinsic curvature
\cite{palatini}. Nevertheless, this dynamical relaxation to the Metric formulation gets spoiled immediately if ${\mathcal{L}}_{\text{m}}\left(\varphi, \Gamma,g\right)$ depends on the affine connection $\Gamma^{\lambda}_{\alpha \beta}$ explicitly. Indeed, in this case variation of the matter action with respect to the connection gives additional contributions to (\ref{relax}) whereby causing $\Gamma^{\lambda}_{\alpha \beta}$
to deviate from $\overline{\Gamma}^{\lambda}_{\alpha \beta}$ as a
function of the matter fields\footnote{The setup of (\ref{action}) is actually not the
most general one. Indeed, the Einstein-Hilbert term can be replaced by a more general
structure containing a generic function of curvature invariants. In this case the deviation
of $\Gamma$ from $\overline{\Gamma}$ can be more involved; in particular, the
equation of motion (\ref{relax}) generalizes to contain new structures
involving derivatives of $\Gamma$ itself.}.

The matter sector consists of all kinds of fields whose interactions are encoded in  ${\mathcal{L}}_{\text{m}}\left(\varphi, \Gamma, g\right)$. In general, at least
at the renormalizable
level, Lagrangians of vector fields do not involve the connection (yet, see \cite{differ2}).
Fermion fields, however, do explicitly depend on $\Gamma$ via the spin connection
in their kinetic terms. Therefore, the equation of motion (\ref{relax}) gets
generically modified by the fermion sector. However, if needed, these contributions can
be compensated by modifying the fermion Lagrangian via contact terms quadratic
in the torsion \cite{differ,torsion}. In this work, following this observation, fermion
contributions to the affinity $\Gamma$ will be discarded. Coming to scalar fields, like
vector fields, they are also independent of the connection when minimally coupled. However,
they develop a direct dependence on the connection, already at the renormalizable level, by
direct coupling to the curvature scalar. Indeed, the Lagrangian of a scalar field $\phi$
\begin{eqnarray}
\label{matter} {\mathcal{L}}_{\text{m}} = \frac{1}{2} g^{\mu \nu}
\partial_{\mu} \phi
\partial_{\nu} \phi - \frac{1}{2} \zeta \phi^2 g^{\mu \nu} \mathbb{R}_{\mu \nu}\left(\Gamma\right) - V(\phi)
\end{eqnarray}
generically involves a nontrivial coupling to the curvature scalar\footnote{Except for Goldstone bosons, such a coupling generically exists for all scalar fields \cite{voloshin}.}
via a dimensionless parameter $\zeta$. This coupling can take any real value\footnote{In the Metric formulation, for $\zeta=1/6$  the action for a scalar field gains local conformal invariance \cite{conformal}. This property does not need to hold in the Palatini formulation where the curvature tensor is intact to transformations on the metric tensor.}.

Given the non-minimally coupled scalar field in (\ref{matter}), the equation of motion
of $\Gamma$ in (\ref{relax}) changes to
\begin{eqnarray}
\nabla^{\Gamma}_{\lambda}\left(
\left({M}^2+\zeta
\phi^2\right) \sqrt{-g} g^{\alpha \beta}\right)=0
\end{eqnarray}
with the solution
\begin{eqnarray}
\Gamma^{\lambda}_{\alpha \beta} = \overline{\Gamma}^{\lambda}_{\alpha \beta}
+ \delta^{\lambda}_{\alpha} \partial_{\beta} \omega(\phi) +
\delta^{\lambda}_{\beta} \partial_{\alpha} \omega(\phi) - g_{\alpha \beta} \partial^{\lambda} \omega(\phi)
\end{eqnarray}
where
\begin{eqnarray}
\label{omega}
\omega\left(\phi\right)=\ln\sqrt{\frac{M^2 + \zeta \phi^2}{M_{\text{Pl}}^2}}
\end{eqnarray}
with the normalization scale $M_{\text{Pl}}$ being
chosen just for convenience\footnote{This new scale must to be introduced here
since $M_{\text{Pl}}$ is the true fundamental scale of gravity in the Einstein frame.}.
It is this very difference of  $\Gamma$ from $\overline{\Gamma}$ that
makes non-minimally coupled scalar fields behave differently in Metric and Palatini formulations.

Another way of seeing differences in scalar field dynamics in Metric and Palatini approaches
comes by changing the frame. Indeed, the action (\ref{action}) with the scalar field Lagrangian
(\ref{matter}) corresponds to the Jordan frame. However, though there is no obvious reason for preferring
one over the other, it is possible to analyze the whole dynamics in the Einstein frame, too. To do this,
it suffices to transform the metric as \cite{trans}
\begin{eqnarray}
\label{mettrans}
g_{\mu \nu} \rightarrow e^{-2 \omega(\phi)}\ g_{\mu \nu}
\end{eqnarray}
after which the coefficient in front of  $g^{\mu\nu} \mathbb{R}_{\mu \nu}\left(\Gamma\right)$ becomes
$\frac{1}{2} M_{\text{Pl}}^2$, as it would be for a minimally-coupled scalar field. This transformation
gives rise to different dynamics for the scalar field in Palatini and Metric approaches since:
\begin{itemize}
\item In the Palatini case $\mathbb{R}_{\mu \nu}\left(\Gamma\right)$ does not depend on the metric and therefore does not change under (\ref{mettrans}).
In the Metric case, the action (\ref{action}) contains $\mathbb{R}_{\mu \nu}\left(\overline{\Gamma}\right)$ and thus the metric from the scratch, and it does change with (\ref{mettrans}).

\item The kinetic term of the scalar field in the Palatini case only gets multiplied by $e^{-2 \omega(\phi)}$ under (\ref{mettrans}). In the Metric case, however, in addition to this rescaling there arises an additional contribution to the kinetic term from the transformation of $\mathbb{R}_{\mu \nu}\left(\overline{\Gamma}\right)$ itself.

\item In the Palatini case, since the coefficient in front of $g^{\mu\nu} \mathbb{R}_{\mu \nu}\left(\Gamma\right)$ is just
$M_{\text{Pl}}^2/2$  the affinity $\Gamma$ reduces to the Levi-Civita one (\ref{levi}). Consequently, the difference between the two formalisms lies in the matter sector wherein (\ref{mettrans}), together with (\ref{omega}), induces different dynamics in Metric and Palatini cases.

\end{itemize}
Several features of the two formalisms mentioned above hold for a generic scalar field theory. Indeed, the
scalar field might be the one that triggers one of those phase transitions in the history of the universe,
might be the inflaton, might be quintessence or might be some other one needed for some specific purpose.
What is important is that the space-time evolutions of the scalar field and the metric are different in the two approaches.

In the next section we provide an explicit case study by analyzing early inflation with a
specific self-interaction potential $V(\phi)$ that can, though not essential
at all for the inflationary regime, also facilitate the electroweak symmetry
breaking in small $\phi$ regime.

\section{Inflation: Metric vs. Palatini Formulations}
The idea of inflation has ever been the most viable framework for
understanding several shortcomings of the standard big bang cosmology \cite{inf}.
During the inflationary epoch the total energy density of the universe
is dominated by that of the vacuum, and the scale factor of the universe grows
exponentially $a(t) \sim e^{ H t}$, $H$ being the Hubble
rate. If this exponential expansion continues for a
time interval $\delta t \sim N/H$ ($N \simgt 60$) then a small
causally connected patch gets magnified sufficiently
to explain the observed flatness, isotropy, and homogeneity
of the universe \cite{inf,CobeNorm}. In spite of these observationally confirmed
advantages, inflationary models suffer from problems
associated with spoiling of the flatness (by quantum corrections)
and super-Planckian values that the inflaton field takes \cite{linde,freese}
(for a recent discussion of the naturalness problem in inflationary models,
see \cite{nima}).

In what follows, we will study inflation in
Metric and Palatini formulations in a comparative fashion, and
confront them with each other as well as with the cosmological
observations. Meanwhile, we will also discuss naturalness problems encountered
in inflationary models \cite{nima} within these two formalisms.
The procedure that will be outlined below is valid for any inflationary
model with its defining inflaton potential $V(\phi)$; however, for concreteness,
we will consider a non-minimally coupled scalar field as in (\ref{matter})
such that its potential $V(\phi)\propto \left(\phi^2 - v^2 \right)^2$ can
facilitate both electroweak symmetry breaking at small $\phi$ and inflation
at large $\phi$. In fact, in \cite{higgsinf} (see also \cite{higgsinf2})
it was shown that one can indeed have a period of early inflation with the
standard model Higgs sector alone by allowing for a non-minimal coupling
between the Higgs field and the curvature scalar
(see also \cite{nonmin} for earlier discussions of inflation with
non-minimal coupling). The analysis below is rather generic and
general yet various results will be presented in parallel
to \cite{higgsinf} so as to allow for quantitative comparison.

The setup of the inflationary scenario we discuss is formed by the
scalar field action (\ref{matter}) replaced in the general
action (\ref{action})
\begin{eqnarray}
\label{main-action}
\mathcal{S} =\int\text{d}^{4}x\sqrt{-g}\left[-\frac{M^{2}+\zeta \phi^2}{2} g^{\mu\nu} \mathbb{R}_{\mu\nu}\left(\Gamma\right)+
\frac{1}{2}g^{\mu\nu} \partial_{\mu}\phi \partial_{\nu}\phi - V(\phi) + \mathcal{L}_{\text{m}}\right]
\end{eqnarray}
where we have separated the rest of the fields plus their interactions with $\phi$ (Yukawa couplings, for instance)
by packing them into $\mathcal{L}_{\text{m}}\equiv \mathcal{L}_{\text{m}}\left(\phi, \varphi, g\right)$. For the self-interaction potential we take
\begin{eqnarray}
V(\phi)=\frac{1}{4} \lambda \left(\phi^{2}-v^{2}\right)^{2}
\end{eqnarray}
which, by construction, facilitates spontaneous symmetry breaking at
$\langle \phi \rangle = v$, and grows like $\phi^{4}$ at large $\phi$.

It is convenient to switch from Jordan frame to Einstein frame in which Newton's constant has its
usual meaning in general relativity. To do this we perform the transformation
of metric in (\ref{mettrans}) where the function $\omega\left(\phi\right)$
is related to the scalar field via relation (\ref{omega}). This transformation
procedure gives
\begin{eqnarray}
\label{met-action}
\mathcal{S}_{\text{M}}&=&\int\text{d}^{4}x \sqrt{-g} \Bigg[-\frac{M_{\text{Pl}}^{2}}{2} g^{\mu \nu} \mathbb{R}_{\mu
\nu}\left(\overline{\Gamma}\right) + \frac{1}{2} \left[e^{-2\omega(\phi)}+
e^{-4\omega(\phi)} \frac{ 6 \zeta^2 \phi^{2}}{M_{\text{Pl}}^{2}}\right] \partial_{\mu}\phi \partial^{\mu}\phi \nonumber\\
&+& e^{-4\omega(\phi)}\left[-V(\phi)+ \mathcal{L}_{\text{m}}\left(\phi, \varphi, e^{-2 \omega(\phi)} g\right) \right]\Bigg]
\end{eqnarray}
in the Metric formalism. One notices that, transformation property of the
Ricci tensor $\mathbb{R}_{\mu \nu}\left(\overline{\Gamma}\right)$ under
(\ref{mettrans}) generates, through non-minimal coupling, an extra
contribution to the kinetic term proportional to $6 \zeta^2 \phi^2$. In the formalism of
\cite{higgsinf} it is this term that largely governs the inflationary expansion.

In the Palatini formulation, on the other hand, the Ricci tensor is invariant under (\ref{mettrans}), and
we find for the corresponding action
\begin{eqnarray}
\label{pal-action}
\mathcal{S}_{\text{P}}&=&\int\text{d}^{4}x \sqrt{-g} \Bigg[- \frac{M_{\text{Pl}}^{2}}{2} g^{\mu \nu} \mathbb{R}_{\mu\nu}\left(\overline{\Gamma}\right) + \frac{1}{2} e^{-2\omega(\phi)} \partial_{\mu}\phi \partial^{\mu}\phi \nonumber\\
&+& e^{-4\omega(\phi)}\left[-V(\phi) + \mathcal{L}_{\text{m}}\left(\phi, \varphi, e^{-2 \omega(\phi)} g\right) \right]\Bigg]
\end{eqnarray}
where we have directly used $\mathbb{R}_{\mu \nu}\left(\overline{\Gamma}\right)$ since $\Gamma = \overline{\Gamma}$
is automatic because $\Gamma$ appears nowhere else (see the discussions leading to (\ref{levi}) in previous section).
Note that varying first the action~(\ref{main-action}) with respect to~$\Gamma$ and subsequently switching to
the Einstein frame via~(\ref{mettrans}) leads to the same result.

A comparative glance at (\ref{met-action}) and (\ref{pal-action}) immediately reveals the difference between the two formalisms:
the kinetic term of the scalar field (now minimally coupled $\phi$ field). This difference influences the definition of the physical scalar field
(to be denoted by $\psi$ hereon) which possesses a canonical kinetic term. Indeed, the physical field $\psi = \psi(\phi)$ follows from
\begin{eqnarray}
\label{h-met}
\frac{\text{d}\psi}{\text{d}\phi}&=&\sqrt{e^{-2\omega(\phi)}+6 M_{\text{Pl}}^{2}(\omega^{\prime}(\phi))^{2}}\nonumber\\
&=&\frac{M_{\text{Pl}}}{M} \sqrt{\frac{1}{1+\zeta\phi^{2}/M^{2}}+\frac{6\zeta^{2}\phi^{2}/M^{2}}{(1+\zeta\phi^{2}/M^{2})^{2}}}
\end{eqnarray}
in the Metric formalism. On the other hand, the physical field in the Palatini formalism derives from
\begin{eqnarray}
\label{h-pal}
\frac{\text{d}\psi}{\text{d}\phi}=\sqrt{e^{-2\omega(\phi)}}=\frac{M_{\text{Pl}}}{M}\sqrt{\frac{1}{1+\zeta\phi^{2}/M^{2}}}
\end{eqnarray}
which differs from (\ref{h-met}) by the absence of the second term in radical sign (which is
generated by the transformation of $\mathbb{R}_{\mu \nu}\left(\overline{\Gamma}\right)$
under (\ref{mettrans})).

In what follows, we will admit large values of non-minimal coupling ($\zeta \gg 1$)
but we will not identify $M$ with $M_{\text{Pl}}$ from the scratch. We will
consider both large and small values of $\psi$ to reveal the potential of the model for
both inflation and electroweak symmetry breaking separately in the Metric and Palatini approaches.

{\it Electroweak regime.}
From (\ref{h-met}) and (\ref{h-pal}) it is clear that
\begin{eqnarray}
\psi\simeq \frac{M_{\text{Pl}}}{M}\, \phi
\end{eqnarray}
in the limit $\phi\sqrt{\zeta}\ll M$. This result is valid for \emph{both} Metric and Palatini
approaches. In this case, the potential of $\psi$ takes the form
\begin{eqnarray}
V(\psi)\simeq\frac{1}{4}\lambda \left(\psi^{2}-\frac{M_{\text{Pl}}^{2}}{M^{2}}v^{2}\right)^{2}
\end{eqnarray}
from which its mass and VEV follow to be
\begin{eqnarray}
m_{\psi}^{2}=2 \lambda \frac{M_{\text{Pl}}^{2} v^{2}}{M^{2}},\,\,\, \langle \psi \rangle =\frac{M_{\text{Pl}} v}{M}\;.
\end{eqnarray}
This setup can trigger the electroweak symmetry breaking once various mass scales are assigned the correct values. There are essentially two possibilities: $(i)$ If the field $\phi$ is the standard model
Higgs field (in unitary gauge) then $v \sim M_{W}$, and thus, $M\sim M_{\text{Pl}}$  for the
minimally-coupled $\psi$ field to keep playing the role of Higgs field like $\phi$. $(ii)$ If only the
field $\psi$, not $\phi$,  is relevant for electroweak symmetry breaking then all one has is the condition
$M_{\text{Pl}} v/M \sim M_{W}$. In this case one can take either $(ii.a)$ $v \sim M_{W}$ and $M\sim M_{\text{Pl}}$
or $(ii.b)$ $v \sim m_{\nu}$ and $M\sim M_W$. Both cases, with $m_{\nu}$ being a typical neutrino mass, lead to the
same result that $\psi$ (not $\phi$) triggers the electroweak symmetry breaking. The case $ii.b)$ is particularly
interesting as it allows $M\sim M_W \ll M_{\text{Pl}}$. It is easy to see that after a suitable rescaling of
the fermion fields the Yukawa couplings in $\mathcal{L}_{\text{m}}\left(\phi, \varphi, g\right)$ imply standard
fermion masses of the order~$m_{\psi}$.

{\it Inflationary regime.} The inflationary regime refers to large values of the fields. More quantitatively, it is achieved when~$\phi\sqrt{\zeta}\gg M$, and in this limit, the Metric formalism yields
\begin{eqnarray}
\phi\simeq\frac{M}{\sqrt{\zeta}}\exp\left(\frac{\psi}{\sqrt{6}M_{\text{Pl}}}\right) \label{eq:Met-phi-h-large}
\end{eqnarray}
provided that $\zeta \gg 1$. This setup agrees with cosmological observations for
\begin{eqnarray}\label{met-M-zeta}
M\simeq M_{\text{Pl}},\,\,\,\zeta\simeq 4.9\times 10^{4} \sqrt{\lambda}
\end{eqnarray}
as has already been derived in \cite{higgsinf}.

We now focus on the Palatini formulation for an analysis of (\ref{pal-action}) for
examining its predictions for the inflationary regime. First, one notes that, unlike
(\ref{h-met}), the condition (\ref{h-pal}) admits direct integration, and hence,
\begin{eqnarray}
\label{phipsi}
\phi=\frac{M}{\sqrt{\zeta}}\sinh\left(\frac{\psi\sqrt{\zeta}}{M_{\text{Pl}}}\right)
\end{eqnarray}
exactly. The main difference from~(\ref{eq:Met-phi-h-large}) is the
appearance of~$\sqrt{\zeta}$ in the argument of the hyperbolic function. Clearly, the
inflationary regime  $\sqrt{\zeta} \phi \gg M$ corresponds to taking $\psi\gg M_{\text{Pl}}/\sqrt{\zeta}$.
Using (\ref{phipsi}) the self-interaction potential of $\psi$ can be computed exactly from (\ref{pal-action}). The
large-field limit of the potential
\begin{eqnarray}
\label{pal-pot}
V(\psi)\simeq\frac{M_{\text{Pl}}^{4}\lambda}{4\zeta^{2}}\left[1-8\left(1+\frac{\zeta v^{2}}{M^{2}}\right)\exp\left(-\frac{2 \psi
\sqrt{\zeta}}{M_{\text{Pl}}}\right)\right]
\end{eqnarray}
is indeed flat, and its flatness and hence relevance for inflation
can be quantified via
\begin{eqnarray}
\epsilon &\equiv& \frac{M_{\text{Pl}}^{2}}{2}\left(\frac{V^{\prime}}{V}\right)^{2}
= 128 \zeta\exp\left(-\frac{4 \psi \sqrt{\zeta}}{M_{\text{Pl}}}\right)\nonumber\\
\eta &\equiv& M_{\text{Pl}}^{2} \frac{V^{\prime\prime}}{V} = -32
\zeta \exp\left(-\frac{2 \psi\sqrt{\zeta}}{M_{\text{Pl}}}\right)
\end{eqnarray}
which are the slow-roll parameters \cite{inf,higgsinf} in the Palatini formulation.

The duration of the inflationary period in units of $H^{-1}$ ($i.e.$
the number of $e$-folds) can be directly computed by using
(\ref{pal-pot}):
\begin{eqnarray}
\label{efold}
N=\frac{1}{M_{\text{Pl}}^{2}}\int_{\psi_{\text{end}}}^{\psi_{\text{start}}}\frac{V}{V'}\,\text{d}\psi\simeq\left.\frac{1}{32\zeta}
\exp\left(\frac{2
\psi\sqrt{\zeta}}{M_{\text{Pl}}}\right)\right|_{\psi_{\text{end}}}^{\psi_{\text{start}}}
\end{eqnarray}
where the inflaton $\psi$ starts with $\psi_{\text{start}}$ and
ends with $\psi_{\text{end}}\simeq
M_{\text{Pl}}\ln(128\zeta)/(4\sqrt{\zeta})$ at which slow-roll
regime is spoiled by $\epsilon \simeq 1$. For
$\psi_{\text{start}}\gg \psi_{\text{end}}$ from (\ref{efold}) one
obtains $32\zeta N=\exp(2
\psi_{\text{start}}\sqrt{\zeta}/M_{\text{Pl}})$. If we further
identify~$\psi_{\text{start}}$ with the field value at which the
COBE scale enters the horizon when~$N\simeq 62$ then the
corresponding normalization condition~\cite{CobeNorm}
\begin{eqnarray}
\frac{V}{\epsilon}\simeq 2 M_{\text{Pl}}^{4}\lambda\zeta^{-1}
N^{2}\simeq (0.027 M_{\text{Pl}})^{4}
\end{eqnarray}
leads to
\begin{eqnarray} \zeta\simeq
1.45\times 10^{10}\, \lambda
\end{eqnarray}
from  which it is clear that the
required value of $\zeta$ in the Palatini approach turns out to be
approximately five orders of magnitude larger than that of the
Metric approach in~(\ref{met-M-zeta}).

At the beginning of the inflationary epoch, the slow-roll parameters at $\psi = \psi_{\text{start}}$ for $N=62$ $e$-folds read as \begin{eqnarray}
\epsilon=\frac{1}{8\zeta N^{2}}\simeq 2.2\times
10^{-15}\,\lambda^{-1},\,\,\,\eta=-\frac{1}{N}\simeq -0.016
\end{eqnarray}
which lead to a spectral index of $n=1-6\epsilon+2\eta\simeq
0.97$, and a tiny tensor to scalar perturbation
ratio~$r=16\epsilon\simeq 10^{-14}$. This very result can be used
to readily falsify the Palatini approach in case a significant
amount of tensor perturbations are observed. Note that higher-order corrections to~$n$ involve the parameters
\begin{eqnarray}
\label{eq:k3k4}
 \xi^2\equiv M^4_\text{Pl}\frac{V^\prime V^{\prime\prime\prime}}{V^2}=\frac{1}{N^2},\,\,\,
 \sigma^3\equiv M^6_\text{Pl}\frac{(V^\prime)^2 V^{(4)}}{V^3}=-\frac{1}{N^3},
\end{eqnarray}
which are significantly smaller than~$\eta$. Therefore, they can be safely neglected in our analysis.

\begin{table}[t]
\begin{center}
\begin{tabular}{|l|l|l|}
\hline
$\phi\sqrt{\zeta}\gg M$ & Metric~\cite{higgsinf} & Palatini \\
\hline
$\phi(\psi)$
& $\frac{M}{\sqrt{\zeta}}\exp{\frac{\psi}{\sqrt{6} M_{\text{Pl}}}}$
& $\frac{M}{\sqrt{\zeta}}\sinh{\frac{\psi\sqrt{\zeta}}{M_{\text{Pl}}}}$ \\
$\zeta$
& $\simeq 4.91\times 10^4\,\sqrt{\lambda}$
& $\simeq 1.45\times 10^{10}\,\lambda$ \\
$\epsilon$
& $\frac{3}{4N^{2}}\simeq 2.0\times 10^{-4}$
& $\frac{1}{8\zeta N^{2}}\simeq 2.2\times 10^{-15}\, \lambda^{-1}$ \\
$\eta$
& $-\frac{1}{N}\simeq -0.016$
& $-\frac{1}{N}\simeq -0.016$ \\
$\frac{\psi_{\text{start}}}{M_{\text{Pl}}}$
& $\frac{\sqrt{6}}{2}\ln{(4N/3)} \simeq 5.4$
& $\frac{1}{2\sqrt{\zeta}}\ln{(32\zeta N)} \simeq 1.3\times 10^{-4}\, \lambda^{-1/2}$ \\
$\frac{\phi_{\text{start}}}{M}$
& $\sqrt{\frac{4N}{3\zeta}}\simeq 4.1\times 10^{-2}\, \lambda^{-1/4}$
& $\sqrt{8N}\simeq 22$ \\
$\frac{\psi_{\text{end}}}{M_{\text{Pl}}}$
& $\frac{\sqrt{6}}{4}\ln{(4/3)} \simeq 0.18$
& $\frac{1}{4 \sqrt{\zeta}}\ln{(128\zeta)} \simeq 5.9\times 10^{-5}\, \lambda^{-1/2}$\\
$\frac{\phi_{\text{end}}}{M}$
& $(\frac{4}{3 \zeta^2})^{1/4} \simeq 4.9\times 10^{-3}\, \lambda^{-1/4}$
& $(\frac{8}{\zeta})^{1/4} \simeq 4.9\times 10^{-3}\, \lambda^{-1/4}$\\
\hline
\end{tabular}
\vspace{0.5cm}
\end{center}
\caption{\label{tab:compare}Main parameters of the inflationary
epoch for $N=62$ $e$-foldings in the Metric and Palatini
formulations.}
\end{table}

For a precise confrontation of the Palatini approach with the Metric
one, we find it useful to compare their predictions for various
quantities in tabular form. We do this in Table \ref{tab:compare}
from which we extract a number of important features:
\begin{itemize}
\item The $\epsilon$ parameter is (in)dependent of $\zeta$ in the (Metric) Palatini approach. In Palatini,
$\epsilon \propto 1/\zeta$, and it gives rise to an extreme
suppression of $\epsilon$ compared to the one in the Metric formalism. One
immediate implication of small $\epsilon$ is that the tensor-to-scalar
ratio of perturbations is approximately $11$ orders of magnitude
smaller in the Palatini approach than in the Metric approach. This is a
testable signature with which one can discard the Palatini approach if
a larger amount of tensor perturbations are found.

\item The $\eta$ parameter is the same (and
much larger than $\epsilon$ in size) for both formalisms; hence,
the spectral index $n$ exhibits almost no change with the formalism,
and equals 0.968 (0.967) for the Palatini (Metric) case. Also the parameters~$\xi^2$ and~$\sigma^3$ in~(\ref{eq:k3k4}) have the same values in both formalisms.

\item  The Einstein frame inflaton $\psi$ (having canonical kinetic
term) starts with $5.4\, M_{\text{Pl}}$ ($1.3\times 10^{-4}\,
M_{\text{Pl}}$) and ends with $0.18\, M_{\text{Pl}}$ ($5.9\times
10^{-5}\, M_{\text{Pl}}$) in the Metric (Palatini) formulation. The
initial value of $5.4\, M_{\text{Pl}}$  is a well known feature of
inflationary models \cite{linde,freese,nima}. It is this
super-Planckian value of $\psi$ that causes a serious naturalness
problem in that all higher--dimensional (Planck-suppressed)
operators become relevant operators. Contrary to this obviously
problematic aspect of the Metric formalism, one immediately
observes that the Palatini approach provides an inflationary epoch
with $\psi$ staying well below the Planck scale. This
sub-Planckian $\psi$--inflaton arises as an important feature of the
Palatini approach. Consequently, this approach provides a
sensible model of inflation since the inflaton does obviously not escape into
the stringy territory.

\item The Jordan frame inflaton $\phi$ (having canonical kinetic
term plus a direct coupling to curvature scalar) starts with $4.1
\times 10^{-2}\, M$ ($22\, M$) in the Metric (Palatini) approach, and
ends with $4.9 \times 10^{-3}\, M$ in both formalisms. In this case,
the Palatini approach seems to require unacceptably large $\phi$
values; however, one notices that the scale $M$, as was discussed
while analyzing electroweak symmetry breaking, does not need to be equal to
$M_{\text{Pl}}$, in fact, it could be as low as $M_W$ without
causing any problem with particle masses. In this sense, the Palatini
approach turns out to give a natural inflationary epoch for both
$\phi$ and $\psi$ inflatons.
\end{itemize}
These itemized features complete the comparative analysis of
inflation in the Metric and Palatini approaches.

\section{Conclusion}
In this work we have performed a comparative analysis of the Metric
and Palatini approaches to gravity. Considering non-minimally
coupled scalar fields, we have discussed how these two formalisms
differ from each other in a general setup in Sec. 2. In Sec. 3 we
have applied findings of Sec. 2 to an inflationary setup whose
fundamental scalar can trigger both electroweak symmetry breaking and
inflation. We have found that both approaches do
have observable signatures of their own. A highly important
feature is that Palatini approach provides a natural inflation
since the inflaton in this formalism stays well below the Planckian
regime. Other features, like strong suppression of tensor
perturbations, form additional distinctive features of the Palatini
approach compared to the Metric one. We thus conclude that a
non-minimally coupled scalar field in the Palatini formulation gives a
sensible inflationary evolution for the early universe.

\section{Acknowledgements}
The work of D.D. was supported by  Alexander von Humboldt-Stiftung
Friedrich Wilhelm Bessel-Forschungspreise and by the Turkish Academy of
Sciences via GEBIP grant.

\end{document}